\documentstyle[12pt,twoside]{article}
\begin{document}
\begin{center}

{\Large\bf 
AN ANALYSIS OF HELIUM\\[5PT]
PRIMORDIAL NUCLEOSYNTHESIS WITH A\\[5PT]
VARIABLE COSMOLOGICAL COUPLING\\[5PT]}
\medskip
 
{\bf F.G. Alvarenga\footnote{e-mail: flavio@cce.ufes.br}
J.C. Fabris\footnote{e-mail: fabris@cce.ufes.br},
S.V.B. Gon\c{c}alves\footnote{e-mail: sergio@cce.ufes.br} and 
J.A.O. Marinho}  \medskip

Departamento de F\'{\i}sica, Universidade Federal do Esp\'{\i}rito Santo, 
CEP29060-900, Vit\'oria, Esp\'{\i}rito Santo, Brazil \medskip

\end{center}
 
\begin{abstract}
The synthesis of helium in the early Universe depends on many input parameters, including the
value of the gravitational coupling during the period when the nucleosynthesis takes place.
We compute the primordial abundance of helium as function of the gravitational coupling,
using a semi-analytical method, in order to track the influence of $G$ in
the primordial nucleosynthesis.
To be specific, we construct a cosmological model with varying $G$, using the Brans-Dicke
theory. The greater the value of $G$ at nucleosynthesis period, the greater the abundance of helium predicted. Using the observational data for the abundance of the primordial helium,
constraints for the time variation of $G$ are established.
\vspace{0.7cm}

PACS number(s): 98.80.Cq, 98.80.Ft
\end{abstract}

\vspace{0.5cm}

\section{Introduction}

The primordial nucleosynthesis is one of the most important achievements of the
cosmological standard model \cite{steigman,malaney}. Using the Einstein's equations, supposing a flat,
isotropic and homogeneous space-time, considering a radiation dominated initial
phase in the evolution of the Universe,
it has been shown that it is possible, assuming initially an equal distribution
of protons and neutrons, to obtain the primordial production of helium,
leading to an abundance of this element of about 24$\%$ of the mass of
the Universe. The observational data indicates an abundance of helium of
$Y_4^{obs} = 0.241\pm0.002$\cite{turner} in mass fraction. Deuterium and lithium,
as well as ${^3}He$, are also produced. However, heavier
elements can not be produced due to the absence of stable nuclei with $A = 5$ and $8$;
heavier nuclei are produced later in the stars. Taking  the ratio of baryons to photons $\eta$, as a free parameter, the predicted abundances agree
with observations with a precision of some percents. For the moment, this is the earliest
test of the standard model, giving confidence that the Universe followed, in general
lines, the evolution predicted by the Hot Big Bang model up to $1$ second of age. It is possible
that the analysis of the anisotropy of cosmic background radiation may lead to tests concerning
earliest moments. But, this remains just a possibility, even
if the results are becoming more and more consistents.
\par
The primordial nucleosynthesis, however,
faces some controverses. As an example, it is considered now that lithium can also be
produced in the stars, and the primordial abundance of this element
could be smaller than it is assumed
today \cite{smith}. Moreover, recent measurements of the second peak of the spectrum of the anisotropy of
the cosmic background radiation, made by the BOOMERANG and MAXIMA atmospheric balloons programs,
indicate a density of baryons in the Universe that agrees
only marginally with that necessary to have the correct primordial production of light elements. In fact the nucleosynthesis requires $\Omega_bh^2 = 0.019\pm0.002$ \cite{olive}, while the measurents
of CMB anisotropy indicates $\Omega_bh^2 = 0.032\pm 0.005$ \cite{jaffe}. However, there are
claims that this discrepancy is less important than indicated by those analysis \cite{netterfield}.
\par
Even if the primordial nucleosynthesis remains one of the most impressive test of
the standard model, it deserves yet more investigations. In particular,
the primordial nucleosynthesis
may be an arena to test the values of some input parameters.
For example, the number of neutrino's
families influence the primordial production of light elements. Hence, the primordial
abundance of light nuclei permits to test
if there are more than three families of leptons and quarks.
\par
In the present work, another effect on the primordial nucleosynthesis will be exploited.
It will
be verified how the value of the gravitational coupling affects the primordial production
of light elements. Hence, the primordial nucleosynthesis will be taken as another way to
verify if the gravitational coupling varies with time. In order to do this, we will work
out a specific case: the Brans-Dicke theory, which is the prototype of a relativistic
gravity theory with varying $G$, will be considered. The Brans-Dicke theory has an interesting
connection with the low energy string action, this being another reason to study it.
In the dust phase of the evolution of the Universe, power law type solution of the Brans-Dicke theory exhibits a decreasing
gravitational coupling. This solution will be matched with the radiative solution, which in its
simplest version coincides with that of the standard model. 
In this way the value of the $G_R$ at the moment of the nucleosynthesis will be connected to its
value today, $G_0$. The value of $G_R$ depends on two input parameters: the ratio of the
densities of radiation and baryonic matter today; the Brans-Dicke coupling parameter
$\omega$. It will be shown that the greater the value of $G$ at the moment of
the nucleosynthesis, the greater the abundance of primordial helium.
\par
The precise calculation of the primordial abundance of light elements is a very
hard task. It implies to use many numerical codes in order to evaluate the
transmutation process involving protons and neutrons to obtain the final ratio between
these nucleons, which determines the final abundance of helium. In order to have very
precise predictions, radiative corrections to the rate of the reactions must be taken
into account.
In the present work we will adopt the semi-analytical method developed in
\cite{bernstein}.
In this method the effects of the Fermi-Dirac statistic are neglected. This leads to
a disagreement with respect to the precise calculation of the order of some percents. Since,
the interest in this work it is to track the influence of a varying $G$ in the
calculation of the primordial abundance of light elements, it is interesting to sacrify
somehow the precison in favour of analytical expressions where the searched effects can
be more easily tracked.
\par
This paper is organized as follows. In section $2$, the Brans-Dicke cosmological model
is developed, including the matching conditions. The relation between $G_R$ and
$G_0$ is established. In section $3$, the computation method presented
in \cite{bernstein} is outlined, stressing the r\^ole played by $G$. In section $4$, the primordial abundance for a
varying $G$ is obtained, in terms of the Brans-Dicke parameter $\omega$. In section
$5$ the conclusions are presented. Some considerations of how to reconcile a
higher value for baryonic density, predicted by CMB anisotropy, with that
one necessary to have abundance of primordial elements in agreement with observations, using a varying $G$ model, are
sketched.

\section{The cosmological scenario}

The Brans-Dicke theory incorporates the space-time variation of the gravitational
coupling in a relativistic theory of gravity \cite{brans}. In order to do so, it couples
a scalar field non-minimally to gravity. Its lagrangian reads,
\begin{equation}
L = \sqrt{-g}\biggr[\phi R - \omega\frac{\phi_{;\rho}\phi^{;\rho}}{\phi}\biggl]
+ L_m
\end{equation}
where $R$ is the Ricci scalar, $\phi$ is a scalar field connected with the
gravitational coupling, $\omega$ is the Brans-Dicke parameter and
$L_m$ is the Lagrangian of matter, which will be taken as a barotropic perfect
fluid with an equation of state $p = \alpha\rho$, $- 1 \leq \alpha \leq 1$. General Relativity,
and consequently a constant gravitational coupling, is recovered when $\omega \rightarrow \infty$. Using the flat Friedmann-Robertson-Walker metric, the equations of motion read,
\begin{eqnarray}
\label{em1}
3\biggr(\frac{\dot a}{a}\biggl)^2 &=& \frac{8\pi\rho}{\phi} + \frac{\omega}{2}\biggr(\frac{\dot\phi}{\phi}\biggl)^2 - 3\frac{\dot a}{a}\frac{\dot\phi}{\phi}
\quad , \\
\label{em2}
\ddot\phi + 3\frac{\dot a}{a}\dot\phi &=& \frac{8\pi}{3+2\omega}(\rho - 3p) \quad ,\\
\label{em3}
\dot\rho + 3\frac{\dot a}{a}(\rho + p) &=& 0 \quad.
\end{eqnarray}
There are two phases which interest us: the dust phase, $\alpha = 0$, and the radiation
phase $\alpha = 1/3$. For these phases, the equations (\ref{em1},\ref{em2},\ref{em3})
admit the following power-law type solutions
\begin{eqnarray}
p &=& 0 \quad \rightarrow \quad a \propto t^{\frac{2 + 2\omega}{4 + 3\omega}} \quad, \quad \phi \propto
t^\frac{2}{4 + 3\omega} \quad , \quad \rho_m \propto a^{-3}\quad ;\\
p &=& \frac{\rho}{3} \quad \rightarrow \quad a \propto t^{1/2} \quad , 
\quad \phi = \mbox{constant} \quad, \quad \rho_r \propto a^{-4} \quad .
\end{eqnarray}
In the above expressions, $\rho_m$ and $\rho_r$ are the density for dust and
radiation respectivelly.
For the radiative phase, the power-law solution coincides with that of the standard model,
implying that the gravitational coupling is constant. But, since during the dust dominated
phase, the gravitational coupling varies with time, the value of the gravitational coupling
at the moment of the nucleosynthesis is, in this model,
different from that normally used. Since
the general features of the Brans-Dicke model for the radiative phase is the same as
in the standard model, we may compute the primordial abundance of light elements with
a minimal modification. More general solutions, of course, can be found. We will discuss
later the general features of these general solutions and their implications for
nucleosynthesis.
\par
In order to extract precise predictions, the solutions for the dust and radiation phases,
as well as their first derivatives,
must be matched. The solutions are rewritten as
\begin{eqnarray}
a = a_0(t - t_m)^\frac{2 + 2\omega}{4 + 3\omega} \quad (p = 0) \quad&,&
\quad \quad a = b_0(t - t_r)^{1/2} \quad (p = \frac{\rho}{3})
\quad ,\\
\rho_m = \frac{\rho_{m0}}{a^3} \quad &,& \quad \rho_r = \frac{\rho_{r0}}{a^4} \quad ,
\end{eqnarray}
where $a_0$, $b_0$, $t_m$, $t_r$ are constants;
$\rho_{m0}$ and $\rho_{r0}$ are the density
for matter and radiation today. Remembering that the gravitational coupling is connected
with the value of the scalar field by $ G = \biggr(\frac{4 + 2\omega}{3 + 2\omega}\biggl)\frac{1}{\phi}$ \cite{weinberg},
and performing the matching of the solutions we find the following relation between the
value of the gravitational coupling today $G_0$ and the value of the gravitational
coupling during the radiative phase $G_R$:
\begin{equation}
G_R = \biggr(\frac{\rho_{m0}}{\rho_{r0}}\biggl)^\frac{1}{1 + \omega}G_0 \quad .
\end{equation}
\par
It is more convenient to work with Planck's unity. Natural unities are employed: 
$c = \hbar = 1$,
$G = 1/\sqrt{M_P}$, $M_P$ being the Planck's mass whose value today is
$M_{P0} = 1.221\times10^{19}GeV$. Hence,
\begin{equation}
\label{MP}
M_{PR} = \biggr(\frac{\rho_{r0}}{\rho_{m0}}\biggl)^\frac{2}{1 + \omega}M_{P0} \quad ,
\end{equation}
where $M_{PR}$ is the value of the Planck's mass at the moment of the nucleosynthesis.

\section{The semi-analytical computation of the primordial nucleosynthesis}

The computation of the nucleosynthesis process in the early Universe involves three
main steps. Initially, at very high energies,
the ratio of neutrons to protons is equal to one. As the Universe expands, the temperature
drops, and reactions involving the neutrons, protons, neutrinos, electrons convert
neutrons into protons. At same time, since the neutrons are free, and unstable,
they decay also into protons. This process continues until the energy is low enough
in order the neutrons to be captured forming deuterium, from which the helium is formed.
The quantity of helium synthetised in this process depends essentially on the quantity
of neutrons that have survived up to the moment they are captured to form deuterium.
The neutrons that are later used to form other elements like lithium are neglected in
the present computation; anyway the others light elements besides helium
represent a very small fraction of the total mass. The detailed analysis
of all these process is quite involved, requiring the use of numerical codes to
evaluate the different transmutation process. However, semi-analytical expressions
can be worked out if the Fermi-Dirac statistics is negletected. This has been done
in \cite{bernstein}, leading to values for the helium abundance with an error of
some percents compared with the precise numerical calculation.
\par
Since all the evaluation of helium abundance 
in the approach used in \cite{bernstein} is very lengthy, we will just summarize the main steps and the relevant quantities. For details, the reader is invited to address
himself to that work.
First we define the ratio of neutrons with respect to the total baryon number:
\begin{equation}
X(T) = \frac{n_n(T)}{n_n(T) + n_p(T)}
\end{equation}
where $n_n(T)$ and $n_p(T)$ are the numbers of neutrons and protons, respectivelly,
as functions of the temperature $T$.
The main equation controling this quantity is
\begin{equation}
\frac{dX(t)}{dt} = \lambda_{pn}(t)(1 - X(t)) - \lambda_{np}(t)X(t) \quad ,
\end{equation}
where $\lambda_{pn}$ and $\lambda_{np}$ are the rates of conversion of protons into
neutrons and neutrons into protons respectively. The main process concerned are
\begin{equation}
\label{trans}
\lambda_{np} = \lambda(\nu + n \rightarrow p + e^-) + 
\lambda(e^+ + n \rightarrow p + \bar\nu) + \lambda(n \rightarrow p + \bar\nu + e^-)
\quad .
\end{equation}
As an example, the first one is given by
\begin{equation}
\lambda(\nu + n \rightarrow p + e^-) = A\int_0^\infty dp_\nu p_\nu^2p_eE_e(1 - f_e)f_\nu
\quad ,
\end{equation}
where $A$ is a coupling constant, the $p$'s denote the momenta of each particle involved in
the process, $E$ the energy and the $f$'s represents the Fermi-Dirac statistics factor.
The last process in (\ref{trans}) represents the neutron decay and it is not considered
in a first evaluation. Later, the final results will be corrected taking it into account.
\par
In \cite{bernstein} the evaluation of the first two rates in (\ref{trans}) is simplified by approaching the
Fermi-Dirac statistics by the Maxwell-Boltzmann one. This is justified by the fact the
temperatures concerned at the moment these process take place are smaller than
the energies of the particles. After a lengthy evaluation of all process, we end up
with the following expression for the neutron abundance factor:
\begin{equation}
X(y) = X_{eq}(y) + \int_0^ydy'e^{y'}X_{eq}^2(y')\exp[K(y) - K(y')] 
\end{equation}
with the following definitions:
\begin{eqnarray}
X_{eq} &=& \frac{1}{1 + e^y} \quad , \\
K(y) &=& b\biggr[\biggr(\frac{4}{y^3} + \frac{3}{y^2} + \frac{1}{y}\biggl)
+ \biggr(\frac{4}{y^3} + \frac{1}{y^2}\biggl)e^{-y}\biggl] \quad , \\
b &=& a\biggr[\frac{45}{4\pi^3N}\biggl]^{1/2}\frac{M_p}{\tau\Delta m^2}
\quad , \quad a = 4A\tau(\Delta m)^5 \quad, \quad y = \frac{\Delta m}{T}
\end{eqnarray}
where $\Delta m$ is the mass difference between neutrons and protons,
$\Delta m = 1.294 MeV$. The fraction of neutrons to baryons at the end of all
those process, $\bar X$, is obtained by making $y \rightarrow \infty$, $\bar X = X(y \rightarrow \infty)$.
\par
As stated before, initially the neutron decay is neglected. To correct the
final results due to
it we must evaluate the time taken in the capture process
forming the deuterium.
This time is given, in principle, by evaluating the capture process until a temperature
of the order of the deuterium binding energy $\epsilon_D \sim 2.225 MeV$.
However, lower energies (of order of $E \sim 0.1 MeV$) must
be considered due to the fact that the enormous number of photons implies that
deuterium dissociation continues to occur even when $T_\gamma < \epsilon_D$. In peforming
this analysis, we
must take into account the evolution of the Universe, which in this case is
reflected by the equation
\begin{equation}
\label{fe}
\frac{\dot T_\gamma}{T_\gamma} = - \biggr[\frac{8\pi\rho}{3M_P^2}\biggl]^{1/2}
\end{equation}
since during the radiative phase $a \propto 1/T_\gamma$, $T_\gamma$ being the
photon temperature, which is approximately equal to the neutrino temperature
at the relevant temperature scales considered in this computation. Moreover,
the energy density is given by
\begin{equation}
\rho = N_{eff}\frac{\pi^2}{30}T^4_\nu \quad ,
\end{equation}
where
\begin{equation}
N_{eff} = N_\nu + \biggr(\frac{11}{4}\biggl)^{4/3}N_\gamma \sim 13 \quad .
\end{equation}
The final results imply that the time of capture is given by
\begin{equation}
t_c = \biggr[\frac{45}{16\pi^3N_{eff}}\biggl]^{1/2}\biggr[\frac{4}{11}\biggl]^{2/3}
\frac{M_P}{T_{\gamma0}^2} + t_0 \quad .
\end{equation}
The constant $t_0$ is
\begin{equation}
t_0 = \frac{11}{6N_{eff}}\biggr[(\frac{11}{4})^{1/3} - 1\biggl]t_1
\quad , \quad t_1 = \biggr[\frac{45}{16\pi^3N_{eff}}\biggl]^{1/2}\biggr(\frac{11}{4}\biggl)^{2/3}\frac{M_P}{T_{\gamma0}^2}
\quad ,
\end{equation}
where $T_{\gamma0}$ is the photon temperature when the neutron capture is accomplished.
The final abundance is given by
\begin{equation}
X_f = \exp{(-t/\tau)}\bar X
\end{equation}
and the final helium abundance by weight is
\begin{equation}
Y_4 = 2X_f \quad .
\end{equation}
Applying all these steps to the standard model, it results $Y_4 \sim 0.243$. Note that this
result is not exactly the same found in \cite{bernstein}, which is $Y_4 \sim 0.247$. We attribute this small discrepancy
to the fact that we used a {\it Mathematica} program instead of a pocket calculator. Moreover,
it is point out in \cite{bernstein} that their results is a kind of upper limit using
the semi-analytical method described there. A more precise
computation, using numerical code, gives $Y_4 \sim 0.241$ \cite{braginsky}.
\par
The above expressions resume briefly the main steps. More important,
it has been shown explicitly where the value of the Planck's mass (consequently, the gravitational
coupling value) appears. This will enable us to compute the variation of the
helium abundance as function of $G_R$, and consequently as function of the Brans-Dicke
parameter $\omega$.
\par
In \cite{bernstein}, it was computed the variation on the helium abundance due
to variations of the chemical potential of the electron neutrino $\mu$, the number of neutrinos $N_\nu$,
the neutron time-life $\tau$ and ratio of photons to baryons $\eta$. They find
\begin{equation}
\Delta Y_4 = - 0.25\mu + 0.014\Delta N_\nu + 0.18\frac{\Delta\tau}{\tau} +
0.009\ln\frac{\eta}{\eta_0}
\end{equation}
where $\eta_0 = 5\times10^{-10}$ is the adopted value for the ratio of baryons to photons.
One of our goals is to add to this expression the term concerning the variation
of the gravitational coupling as a perturbation around its value computed using the
value of
$G$ today.

\section{Helium abundance with varying G}

Now we turn to the computation of the helium abundance with the cosmological model
developed in section $2$. First of all, we notice that since in the radiative phase
the scalar field is constant, all development exhibits in the previous section is valid;
we must just to compute the value of the gravitational coupling (Planck's mass) in view
of the fact that in the later dust phase the gravitational coupling varies with time.
We fix in (\ref{MP}) $\rho_{m0} \sim 10^{-29}g/cm^3$ and $\rho_{r0} \sim 0.950\times10^{-33}g/cm^3$,
which are approximately the densities of matter and radiation today \cite{misner}. Using
$M_{P0} \sim 1.221\times10^{19}Gev$, we can then compute the values of $M_{PR}$ at
the radiative phase in terms of $\omega$. With the value of $M_{PR}$, we can
compute the helium mass fraction. The values are the following:
\begin{center}
\begin{tabular}{||r||r||r||}
\hline\hline
{\bf $\omega$}&{\bf $M_{PR}$} (in $GeV$)&{\bf $Y_4$}\\ \hline\hline
$- 1$&$\infty$&$1$ \\ \hline
$- 0.5$&$1.160\times10^{15}$&$0.961$ \\ \hline
$0$&$1.190\times10^{17}$&$0.818$ \\ \hline
$1$&$1.205\times10^{18}$&$0.612$ \\ \hline
$10$&$8.015\times10^{18}$&$0.318$ \\ \hline
$50$&$1.115\times10^{19}$&$0.259$ \\ \hline
$100$&$1.166\times10^{19}$&$0.251$ \\ \hline
$500$&$1.210\times10^{19}$&$0.244$ \\ \hline
$1,000$&$1.215\times10^{19}$&$0.243$ \\ \hline
$10,000$&$1.220\times10^{19}$&$0.243$ \\ \hline
\end{tabular}
\end{center}
\par
From this table, we can see that all matter becomes composed of helium for
$\omega = - 1$ (the case where the Brans-Dicke cosmology coincides with
the string cosmology at low energy level \cite{gasperini}), and it approaches the asymptotic value of the standard model
as $\omega$ increases. This result is easy to understand. Since in the radiative
phase the Friedmann equation (\ref{fe}) is valid, as $G$ increases the rate of
the expansion of the Universe also increases. Hence, the temperature drops very
quickly and the duration of the nucleosynthesis era becomes shorter. All effects
described in the preceding section contribute to decrease the quantity of neutrons.
If the duration of the nucleosynthesis era becomes smaller, than more neutrons survive
to be captured in deuterium, forming helium later. When $\omega = - 1$ the value of the
gravitational coupling diverges, there is an instantaneous transition from the
radiative phase to the dust phase, and all initial neutrons survive.
\par
We may evaluate now how a small change in the value of the gravitational coupling
affects the nucleosynthesis as an approximation to the
standard model. This will permit us to establish constraints on the
variation of $G$ by using the nucleosynthesis observational results.
The value of $G$ affects essentially the parameters $b$, $t_c$ and $T_{\gamma0}$.
Writing $G = G_0 + \Delta G$, ($M_P = M_{P0} + \Delta M_P$), and introducing this quantity
in the computation steps described before, we find that the helium matter fraction varies as
\begin{equation}
\Delta Y_4 = 0.088\frac{\Delta G}{G_0} \quad .
\end{equation}
It is generally argued that the nucleosynthesis observational results coincides with
the theoretical results by with a precision of $1-2\%$. Hence
\begin{equation}
\frac{\Delta G}{G_0} \sim 1 \rightarrow \frac{\dot G}{G} \sim 10^{-10} \mbox{years}^{-1} 
\end{equation}
where we have, in the last step, divided the first expression by the age of the Universe,
assumed to be $t_U \sim 2\times10^{10}$ years,
to obtain an estimation of the fractional time variation of $G$. This agrees in order of
magnitude with
the estimations of the time variation of $G$ using other methods\cite{shapiro,braginsky,will}.
Of course, a more precise comparison with other experimental determinations
of $\frac{\dot G}{G}$
is not possible due to the approximations made.

\section{Conclusions}

Primordial nucleosynthesis is considered as one of the most precise tests of the
standard cosmological scenario. With it, the abundances of helium, deuterium and lythium,
due to the primordial processes,
are computed. Comparison with observations showed that the predicted abundances agree
with the observed ones by some percents. This is an impressive result since the primordial
nucleosynthesis ocurred in the first seconds of the existence of the Universe.
This high precise cosmological test may be used to constrain the values of some input
parameters, like the number of neutrinos and the ratio of baryons to photons.
\par
In the present work, we have estimated the influence of a possible variation of the
gravitational coupling $G$ in the predicted abundance of light elements. In order to
be specific, a Brans-Dicke cosmological scenario was constructed, with a radiative phase
identical to that of the standard model, followed by a dust phase. The gravitational
coupling decreases during the dust phase, while it remains constant during the radiative
phase. Hence, the value of $G$ at the moment of the nucleosynthesis is, in this model,
higher than the usual one.
\par
The results indicate that for a larger $G$, the abundance of helium is higher. This is
due to the fact that all the process occuring during the nucleosynthesis contribute
to decrease the number of neutrons. A larger value of $G$ leads to a faster expansion
of the Universe, decreasing the duration of the nucleosynthesis era. Consequently, the
final number of neutrons that will survive, forming later the helium, is higher.
The precision of the measurements of the abundance of the primordial elements permits
then to estimate the possible variation of $G$. We found however, that window of
allowed values is not larger than that one obtained by another kind of experiment.
\par
In all computation, it was used the simplest power-law type solution of the Brans-Dicke
theory for the radiative phase, which is identitical to the standard model one. This simplify a
lot the evaluation of the helium abundance. However, it is possible in principle to
consider more complicated scenarios. In fact, for the radiative phase, the Brans-Dicke
scalar-field can be written as
\begin{equation}
\dot\phi = \frac{C}{a^3} \quad ,
\end{equation}
where $C$ is a constant. If $C$ is positive, the gravitational coupling decreases with
time during the radiative phase; if it is negative, the gravitational
coupling increases with time. The case studied before corresponds to $C = 0$.
For $C \neq 0$, we have the following solution for the scale factor:
\begin{equation}
a(\xi) = a_0\biggr[\xi - B\biggl]^{\frac{1 - \sqrt{\frac{3}{8\bar\omega}}}{2}}\biggr[\xi + B\biggl]^{\frac{1 + \sqrt{\frac{3}{8\bar\omega}}}{2}} \quad ,
\end{equation}
where $B = \sqrt{\frac{\bar\omega}{6}}C$, $\bar\omega = \omega + \frac{3}{2}$ and $\xi$ is
the conformal time
\begin{equation}
\xi = \int\frac{dt}{a} \quad .
\end{equation}
It will be interesting to consider such more general model, including a possible increase
of the gravitational coupling during the radiative era, since it can be
a possible solution to the discrepancy of the baryonic density obtained from
nucleosynthesis and from CMB anisotropy. This leads, however, to more
important modifications in the preceding computation, since some basic equations, like
(\ref{fe}) must be modified by including the contribution of the
non-minimal coupled scalar field. We hope to present this more general analysis in the
future.

{\bf Acknowledgements:} We thank CNPq (Brazil) for partial financial support.


\begin{thebibliography}{100}
\bibitem{steigman} A.M. Boesgaard and G. Steigman, Ann. Rev. Astron. Astrophys. {\bf 23}, 319(1985);
\bibitem{malaney} R.A. Malaney and G.J. Mathews, Phys. Rep., {\bf 229}, 145(1993);
\bibitem{turner} S. Buerles, K. M. Nollett and M.S. Turner, {\it Big-bang nucleosynthesis predictions for precision cosmology}, astro-ph/0010171;
\bibitem{smith} V.V. Smith and D.L. Lambert, Astrophys. J. {\bf345}, L75(1989);
\bibitem{olive} K.A. Olive, Phys. Rep. {\bf 333}, 389(2000);
\bibitem{jaffe} A.H. Jaffe et al., Phys. Rev. Lett. {\bf 86}, 3475(2000);
\bibitem{netterfield} C.B. Netterfield et al., {\it A measurement by BOOMERANG of multiple
peaks in the angular power spectrum of the cosmic microwave background}, astro-ph/0104460;
\bibitem{bernstein} J. Bernstein, L.S. Brown and G. Feinberg, Rev. Mod. Phys. {\bf 61}, 25(1989);
\bibitem{brans} C. Brans and R.H. Dicke, Phys. Rev. {\bf 124}, 925(1961);
\bibitem{weinberg} S. Weinberg, {\bf Gravitation and Cosmology}, Wiley, New York(1972);
\bibitem{turner1} R.E. Lopez and M.S. Turner, Phys. Rev. {\bf D59}, 3502(1999);
\bibitem{shapiro} I.I. Shapiro et al. Phys. Rev. Lett. {\bf 26}, 27(1971);
\bibitem{braginsky} V.B. Braginsky, C.M. Caves and K.S. Thorne, Phys. Rev. {\bf D15}, 2047(1977);
\bibitem{misner} C.W. Misner, K.S. Thorne and J.A. Wheeler, {\bf Gravitation},
W.H. Freeman, San Francisco(1973);
\bibitem{gasperini} M. Gasperini, {\it Elementary introduction to pre-big bang cosmology
and to the relic graviton background}, hep-th/9907067;
\bibitem{will} C.M. Will, {\bf Theory and experiment in gravitational physics},
Cambridge University Press, Cambridge(1993).
\end{thebibliography}
\end{document}